\documentclass[aps,prx,twocolumn,superscriptaddress,a4paper,english,longbibliography]{revtex4-2}

\usepackage{times}
\usepackage{color}
\usepackage[colorlinks=true,urlcolor=blue,citecolor=blue]{hyperref}
\usepackage{url}
\usepackage{soul}
\usepackage{graphicx}
\usepackage{amsmath}
\usepackage{subfigure}
\usepackage{xcolor}
\usepackage{amssymb}
\usepackage{latexsym}
\usepackage{hyperref}
\DeclareGraphicsExtensions{.png,.eps}
\usepackage{float}
\usepackage{ulem}
\usepackage{appendix}
\usepackage{etoolbox}

\usepackage{xcolor}
\usepackage[urlcolor=blue]{hyperref}      
\hypersetup{
    colorlinks = true,                    
    citecolor = {blue},
    linkcolor = {purple},
           }
\begin{document}	
	
\title{Unsupervised Recognition of Informative Features via Tensor Network Machine Learning and Quantum Entanglement Variations}

\author{Sheng-Chen Bai}
\thanks{These two authors have contributed equally to this work.}
\affiliation{Department of Physics, Capital Normal University, Beijing 100048, China}

\author{Yi-Cheng Tang}
\thanks{These two authors have contributed equally to this work.}
\affiliation{Department of Physics, Capital Normal University, Beijing 100048, China}

\author{Shi-Ju Ran}
\email[Corresponding author. Email: ]{sjran@cnu.edu.cn}
\affiliation{Department of Physics, Capital Normal University, Beijing 100048, China}
\date{\today}

\begin{abstract}
	Given an image of a white shoe drawn on a blackboard, how are the white pixels deemed (say by human minds) to be informative for recognizing the shoe without any labeling information on the pixels?  Here we investigate such a ``white shoe'' recognition problem from the perspective of tensor network (TN) machine learning and quantum entanglement. Utilizing a generative TN that captures the probability distribution of the features as quantum amplitudes, we propose an unsupervised recognition scheme of informative features with the variations of entanglement entropy (EE) caused by designed measurements. In this way, a given sample, where the values of its features are statistically meaningless, is mapped to the variations of EE that statistically characterize the gain of information. We show that the EE variations identify the features that are critical to recognize this specific sample, and the EE itself reveals the information distribution of the probabilities represented by the TN model. The signs of the variations further reveal the entanglement structures among the features. We test the validity of our scheme on a toy dataset of strip images, the MNIST dataset of hand-drawn digits, the fashion-MNIST dataset of the pictures of fashion articles, and the images of brain cells. Our scheme opens the avenue to the quantum-inspired and interpreted unsupervised learning, which can be applied to, e.g., image segmentation and object detection. 
\end{abstract}
\maketitle

\section{Introduction}

Machine learning (ML) such as deep learning have gained tremendous successes in an extremely wide range of fields such as computer vision and natural language processing. Such methods have strong demands on the labeled samples to extract useful information in a data-driven manner. However, labeled data are rare in many scenarios such as the scientific images. Exploring efficient and reliable schemes for the unsupervised~\cite{B89uML} and few-shot~\cite{WYKN21FSML} learning is at the cutting edges of ML and artificial intelligence.

A promising pathway to the unsupervised learning is to develop interpretable ``white-box'' ML schemes~\cite{BGYG+18MLinterp} by cooperating with the probabilistic theories and models, where we have for instance the information bottleneck theory~\cite{GGG03infoML} and Bayesian inference~\cite{LFP03BayesML}. In recent years, tensor network (TN), which originated from the fields of quantum physics~\cite{VMC08MPSPEPSRev, CV09TNSRev, RTPC+17TNrev, O19TNrev, CPDSV21TNSrev}, sheds light on the novel quantum ML schemes~\cite{BWPR+17QML} interpreted by quantum probabilistic theories and quantum many-body physics. TN ML has been successfully applied to the supervised, unsupervised, and reinforcement learning for various tasks including classification~\cite{SS16TNML, LRWP+17MLTN, SPLRS19GTNC, CWZ21PEPSML}, generation~\cite{HWFWZ17MPSML, CWXZ19generateTTNML}, feature selection~\cite{LZLR18entTNML}, compressed sampling~\cite{RSF+20TNCS}, anomaly detection~\cite{wang2020anomaly}, and etc. Experiments of running TN ML on quantum hardware are also in hot debate~\cite{WXYRX21MPSphoto, WAQ21GTNexp}.


\begin{figure}[tbp]
    	\centering
    	\includegraphics[angle=0,width=1\linewidth]{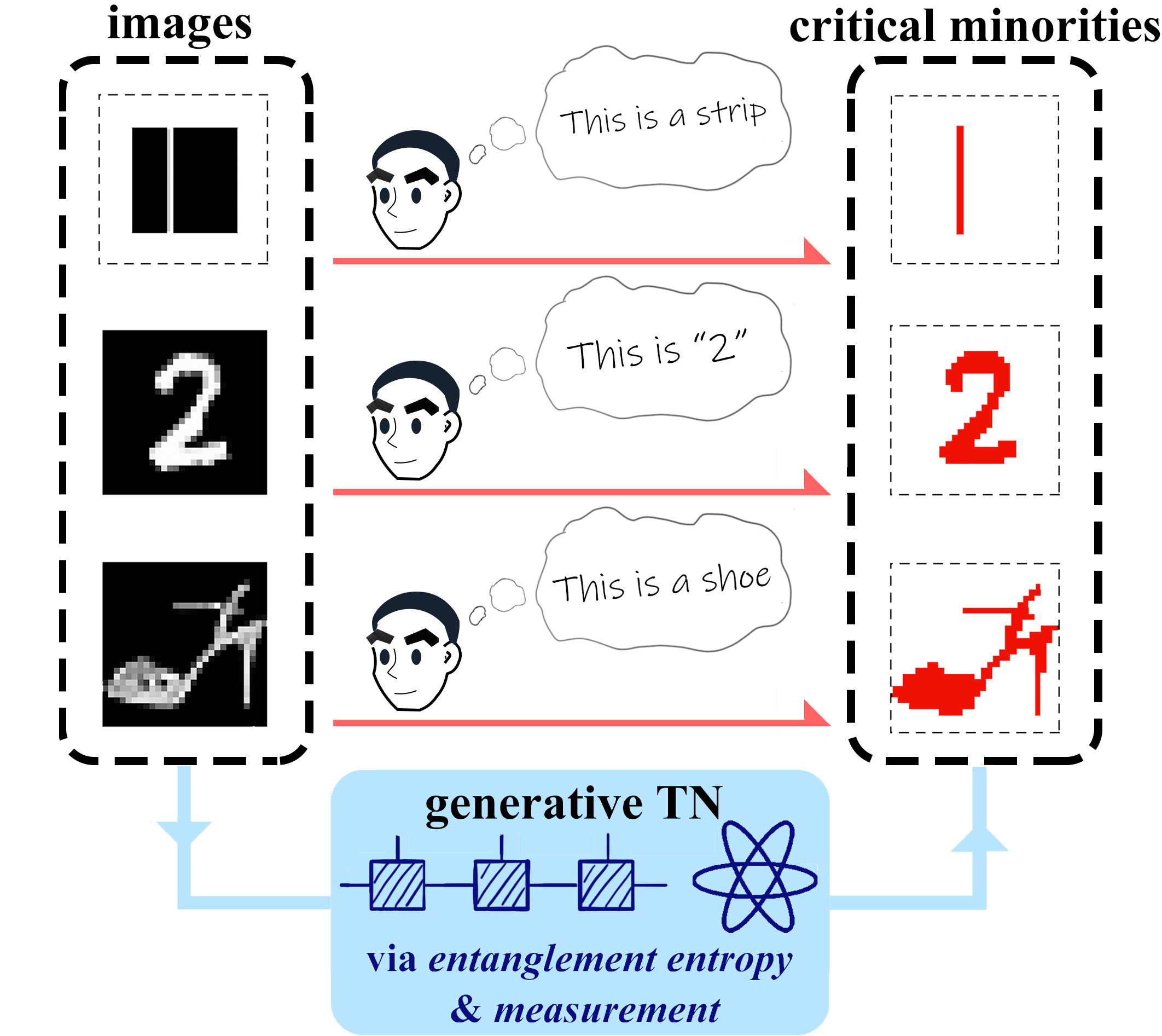}
    	\caption{(Color online) Provided with the images of a vertical strip, a digit ``2'' and a shoe drawn on the black background (left-hand-side), a human mind can easily recognize the shapes of the images as the white pixels, which are dubbed as the critical minorities (right-hand-side), without explicitly learning any labeling information on the pixels. We propose an unsupervised TN scheme to identify the critical minority by the variations of the entanglement entropies.}
    	\label{fig-idea}
\end{figure}

In this work, we propose to unsupervisedly recognize the informative features via the generative TN~\cite{HWFWZ17MPSML} and its entanglement entropy (EE)~\cite{LZLR18entTNML}.  Given an image, a ML model simply sees a bunch of numbers (pixels) that are statistically meaningless. However, a human mind can easily recognize the critical pixels for identifying the content of the image without explicitly learning any labeling information on the pixels. Taking the three images on the left-hand-side of Fig.~\ref{fig-idea} as examples, a human mind could easily recognize the white pixels picturing the objects (strip, ``2'', and shoe). To seek for a mathematical understanding and modeling of such recognition,  we suggest to map the pixels of a given image to the statistically meaningful quantities by the single-qubit measurements on a generative TN according to the pixels of this image. Specifically, we propose to use the average variations of EE (denoted as $\langle \delta S \rangle_{m'}$ for the $m'$-th pixe) for unsupervised feature selection. The pixels with large $\langle \delta S \rangle_{m'}$ (dubbed as the critical minority) outline the critical shape for recognizing this image, coinciding with the human minds. We test the proposed method on a toy dataset of strips, the MNIST~\cite{MNIST_web} and fashion-MNIST~\cite{fMNIST} datasets. Our scheme differs from the existing unsupervised feature selection methods~\cite{SCM20UFSrev}, such as the filter methods (see, e.g., Refs.~[\onlinecite{10.1093/bioinformatics/btl214, TABAKHI20151024}]), the clustering and dimensionality reduction methods (e.g., Refs.~[\onlinecite{10.5555/1005332.1016787, KIM20115704, YAO2015100}]), and etc. These methods require multiple samples and their processings. We finally apply our method for image segmentation provided with just one image of brain cells~\cite{segCell} without any labeling information, and raise the open question on generalizing to the feature selection by multi-qubit measurements.

\section{Mapping to the variations of entanglement entropies via generative tensor network and measurements}

The first step of modeling the probability distribution by a generative TN is to map the samples to the quantum Hilbert space (known as the feature map~\cite{SS16TNML}) as
\begin{eqnarray}
 \boldsymbol{v}^{[n]} = \prod_{\otimes m=1}^M \left[\cos\left(\frac{x^{[n]}_m\pi}{4}\right), \sin\left(\frac{x^{[n]}_m\pi}{4}\right) \right]^{T},
 \label{eq-featuremap}
\end{eqnarray}
with $\boldsymbol{x}^{[n]} = [x^{[n]}_{1}, \ldots, x^{[n]}_{M}]$ the $n$-th sample consisting of $M$ features~\cite{note1}. One can see that $\boldsymbol{v}^{[n]}$ is an $M$-th order tensor or a $2^{M}$-dimensional vector. Obviously, $\boldsymbol{v}^{[n]}$ is normalized satisfying $\left| \boldsymbol{v}^{[n]} \right|=1$ (L2 norm), and can be considered as the coefficient vector of an $M$-qubit quantum product state.

Considering the generative model as a normalized $M$-th order tensor $\boldsymbol{\Psi}$, the probability of generating a specific sample $\tilde{\boldsymbol{x}} = (\tilde{x}_{1}, \ldots, \tilde{x}_{M})$ follows Born's probabilistic interpretation of quantum mechanics with
\begin{eqnarray}
P(\tilde{\boldsymbol{x}}) = \left| \sum_{s_{1} \ldots s_{M}}  \Psi_{s_{1} \ldots s_{M}}  \tilde{v}_{s_{1} \ldots s_{M}} \right|^{2},
 \label{eq-py}
\end{eqnarray}
with $\tilde{\boldsymbol{v}}$ defined by Eq.~(\ref{eq-featuremap}) with $\tilde{\boldsymbol{x}}$. $\boldsymbol{\Psi}$ can be regarded in general as an $M$-qubit entangled state. A generative TN is trained so that the probability of generating each sample in the training set approaches to $1/N$ with $N$ the total number of training samples. To this end, Han \textit{et al}~\cite{HWFWZ17MPSML} proposed to write $\boldsymbol{\Psi}$ into a widely-used TN, namely matrix product state (MPS~\cite{PVWC07MPSRev, VMC08MPSPEPSRev, RTPC+17TNrev}, also known as the tensor-train form~\cite{O11TTD}), which is formed by $M$ local tensors $\{\boldsymbol{A}^{[m]}\}$ as
\begin{eqnarray}
\Psi_{s_{1} \ldots s_{M}} =  \sum_{a_{1} \ldots a_{M-1}} && A^{[1]}_{s_{1} a_{1}} A^{[2]}_{s_{2} a_{1}a_{2}} \ldots \nonumber \\ && A^{[M-1]}_{s_{M-1} a_{M-2}a_{M-1}} A^{[M]}_{s_{M} a_{M-1}}.
 \label{eq-MPS}
\end{eqnarray}
The indexes $\{s_{m}\}$ are called the physical indexes, and $\{a_{m}\}$ the virtual indexes whose dimension (denoted as $\chi$) is a hyper-parameter that controls the parameter complexity of the MPS. A sweep algorithm~\cite{SS16TNML} inspired by the density matrix renormalization group~\cite{W92DMRG, W93DMRG} is used to optimize the local tensors $\{\boldsymbol{A}^{[m]}\}$ to minimize the following loss
\begin{eqnarray}
L = -\frac{1}{N} \sum_{n=1}^{N} \ln{P(\boldsymbol{x}^{[n]})}.
 \label{eq-NLL}
\end{eqnarray}

\begin{figure}[tbp]
    	\centering
    	\includegraphics[angle=0,width=1\linewidth]{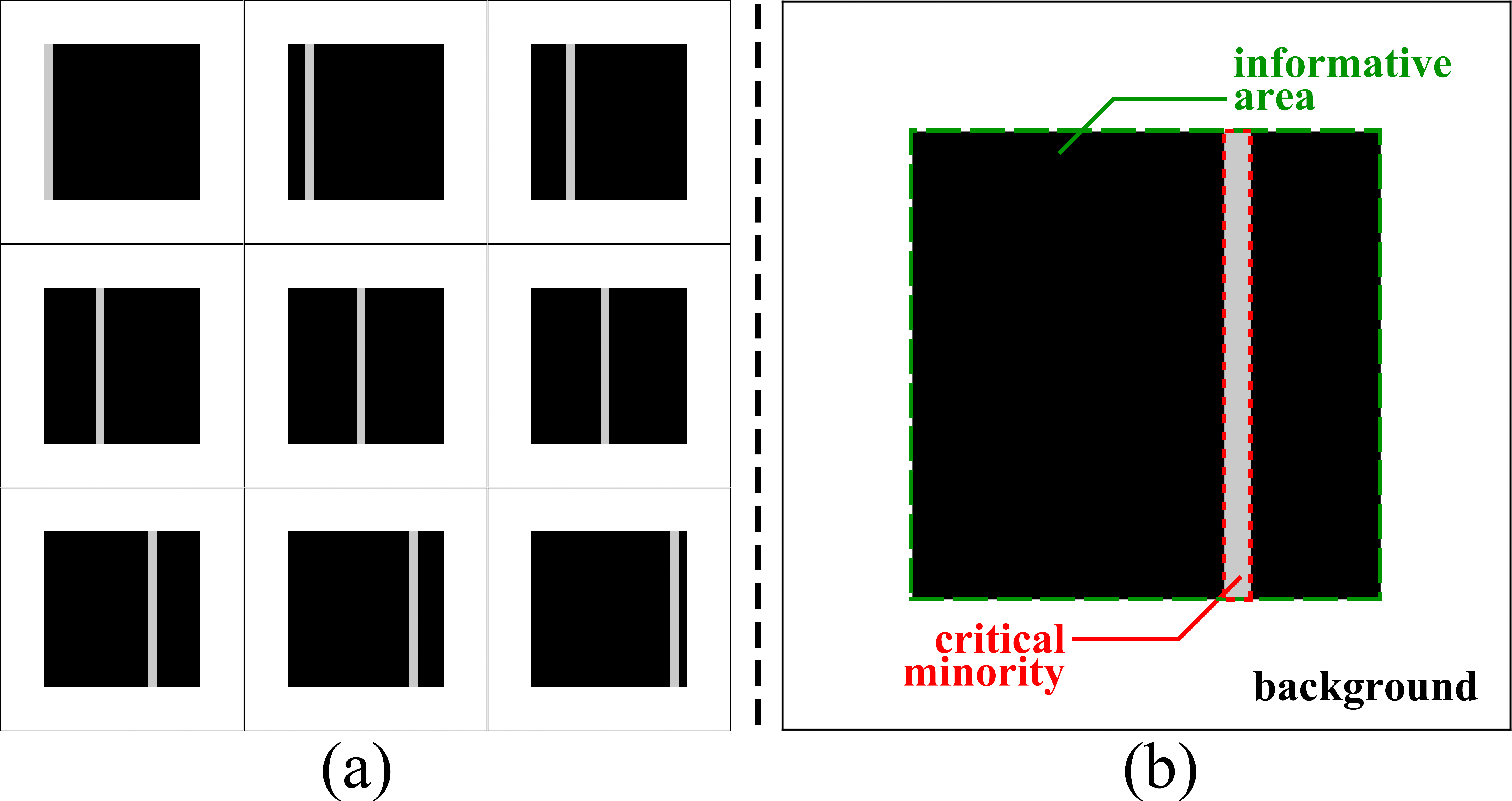}
    	\caption{(Color online) (a) Several samples in a toy dataset of vertical strips, and (b) the illustrations of the background, informative area, and critical minority.}
    	\label{fig-stripe}
\end{figure}

To explain the main idea of this work, we design a toy dataset of vertical strips [Fig.~\ref{fig-stripe} (a)]. The whole region consists of three parts [Fig.~\ref{fig-stripe} (b)]. In the outer rim, the pixels are taken to be white (with $x_m=0$) for all samples thus contain no information at all, which we name as the background. A vertical strip (with $x_m=0.1$) appears at different positions in the black square region (with $x_m=1$) in the middle, which we dub as the informative area. Particularly, the pixels of the strip in each image are referred as the critical minority that we assume to carry the most critical information of the image. This is a reasonable assumption as a human, after briefly reading the images in this dataset, could easily recognize the ``moving'' strips as the critical minority.

\begin{figure}[tbp]
    	\centering
    	\includegraphics[angle=0,width=1\linewidth]{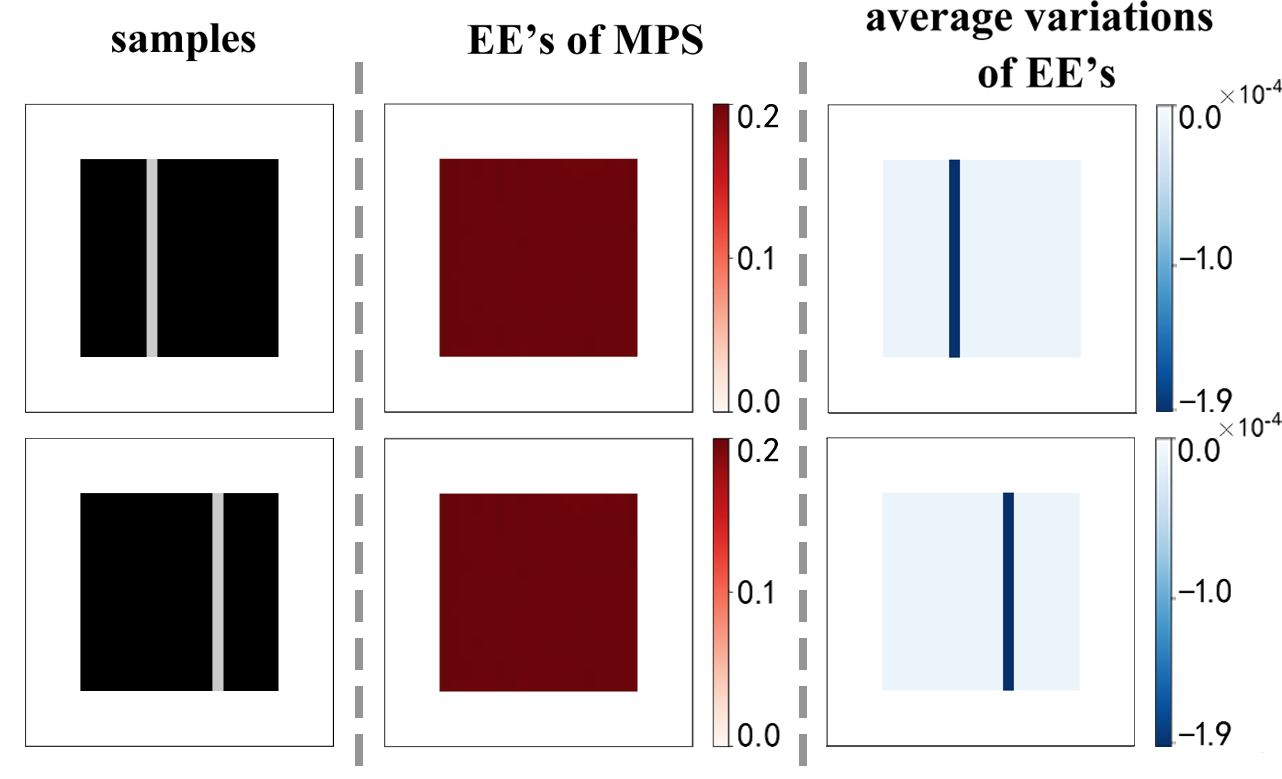}
    	\caption{(Color online)We take two images (first column) from the dataset of strips, and show the EE of the generative MPS [second column; see Eq.~(\ref{eq-EE})] and the average variations of EE [third column; see Eq.~(\ref{eq-dEE})] by measuring according to the images.}
    	\label{fig-stripes_data}
\end{figure}

The main points of this work are as follows: 
\begin{itemize}
	\item The EE of the MPS indicate the informative area and the background for the dataset (which is in general sample-independent, similar to the existing unsupervised feature selection methods~\cite{SCM20UFSrev});
	\item The average variations of the EE by measuring the MPS indicate the critical minority of a specific image (which is sample-dependent).
\end{itemize}
Fig.~\ref{fig-stripes_data} take two samples (see the first column) as the examples to demonstrate the entanglement information obtained from the generative MPS. The second column shows the EE of MPS. The EE corresponding the $m$-th pixel (or physical index) is defined as
\begin{eqnarray}
 S_{m} = -\text{Tr}\left( \boldsymbol{\rho}^{[m]} \ln{\boldsymbol{\rho}^{[m]}} \right),
 \label{eq-EE}
\end{eqnarray}
where $\rho^{[m]}_{s_{m} s'_{m}} = \sum_{/m} \Psi^{\ast}_{s_{1} \ldots s_{m} \ldots s_{M}} \Psi_{s_{1} \ldots s'_{m} \ldots s_{M}}$ is the reduced matrix of the $m$-th physical index ($\sum_{/m}$ means to sum over all but the $m$-th physical indexes). The EE of a single qubit characterizes the amount of uncertainty that can be reduced by measuring this qubit. Thus, a larger EE indicates more information obtained by measuring this qubit, and \textit{vice versa}. In the background, the EE is zero as expected since mathematically the corresponding qubits form the unentangled product states for the strip dataset. The EE in the informative area is approximately uniform with $S_{m} \simeq 0.2$, since the probabilities of having a strip at different positions are uniform. The distribution of the EE clearly identifies the pixels that contain non-trivial information.

To identify the informative features that are critical for a specific sample, we investigate the variations of the EE by measuring on one qubit according to the value of the corresponding feature. Given a specific sample $\tilde{\boldsymbol{x}}$, we measure on the $m'$-th qubit and have
\begin{eqnarray}
 \Phi^{[m']}_{s_{1} \ldots s_{m'-1} s_{m'+1} \ldots s_{M}} = \sum_{s_{m'}} v^{[m']}_{s_{m'}}  \Psi_{s_{1} \ldots s_{M}},
 \label{eq-Phik}
\end{eqnarray}
with the vector $\boldsymbol{v}^{[m']} $ obtained using the feature map on the $m'$-th feature $\tilde{x}_{m'}$ of the given image. By normalizing $\boldsymbol{\Phi}^{[m']} / |\boldsymbol{\Phi}^{[m']}| \to \boldsymbol{\Phi}^{[m']}$, it represents an ($M-1$)-qubit state that captures the posterior probability distribution of the ($M-1$) unmeasured features in the condition of knowing $\tilde{x}_{m'}$. The average variation of the EE after the measurement is defined as 
\begin{eqnarray}
 \langle \delta S \rangle_{m'} =\frac {\sum_{m \neq m'} \left( S'_{m}-S_{m} \right)} {M-1},
 \label{eq-dEE}
\end{eqnarray}
with $S_{m}$ and $S'_{m}$ the EE of the $m$-th qubit before and after the measurement, respectively. In short, Eq.~(\ref{eq-dEE}) maps a given sample of $M$ features to $\langle \delta S \rangle_{m'}$ ($m'=1, \ldots, M$). 

The third column of Fig.~\ref{fig-stripes_data} shows the $\langle \delta S \rangle_{m'}$ ($m' = 0, \ldots, M$) by measuring each qubit respectively according to the image given in the first column. For the critical minority (the pixels of the white strip), we clearly obtain much larger EE variations with $\langle \delta S \rangle_{m'} \sim O(10^{-4})$. In the informative area but outside the strip, we have $\langle \delta S \rangle_{m'} \sim O(10^{-5})$, which are non-zero but much smaller than those of the critical minority. For the background, we have $\langle \delta S \rangle_{m'} = 0$ since the EE's before and after the measurement are zero. Our results show that $\langle \delta S \rangle_{m'}$ can mark the critical minority fairly well, though we do not have any prior information on labeling the pixels. The larger EE variations in the strip are essentially due to the fact that the black pixels in the strip are a monitory compared with the rest ones within the informative area. Consequently, knowing a pixel to be white (in the informative area) will largely reduce the EE of the qubits in the same column by knowing the position of the strip. In comparison, known a pixel to be black only excludes this column as the position of the strip, where the decrease of the EE should be relatively small. The generative MPS captures such properties in a simple manner: the qubits are more entangled strip-wisely. 

To provide an intuitive understanding, let us consider the following three-qubit state as a simplest example
\begin{eqnarray}
  \boldsymbol{\Psi} = \frac{1}{\sqrt{2}} 
\left[
\begin{array}{cc}  
1 \\ 0
\end{array} 
\right]
\otimes
\left(
  \left[
\begin{array}{cc}  
1 \\ 0
\end{array}
\right]  \otimes
\left[
\begin{array}{cc}  
0 \\ 1
\end{array}
\right] +
\left[
\begin{array}{cc}  
0 \\ 1
\end{array}
\right]  \otimes
\left[
\begin{array}{cc} 
1 \\ 0
\end{array}
\right] \right).
  \label{eq-singlet}
\end{eqnarray}
This state can be considered to describe the probability distribution of two samples $\boldsymbol{x}^{[1]} = (0, 0, 1)$ and $\boldsymbol{x}^{[2]} = (0, 1, 0)$, with $P(\boldsymbol{x}^{[1]}) = P(\boldsymbol{x}^{[2]}) = 0.5$. The last two qubits form a maximally entangled ``singlet'' state.  From Eq.~(\ref{eq-EE}), we have $S=0$ for the first qubit, and $S=\ln{2}$ for the last two. The first qubit can be recognized as the background. 

 \begin{figure}[tbp]
    	\centering
    	\includegraphics[angle=0,width=0.9\linewidth]{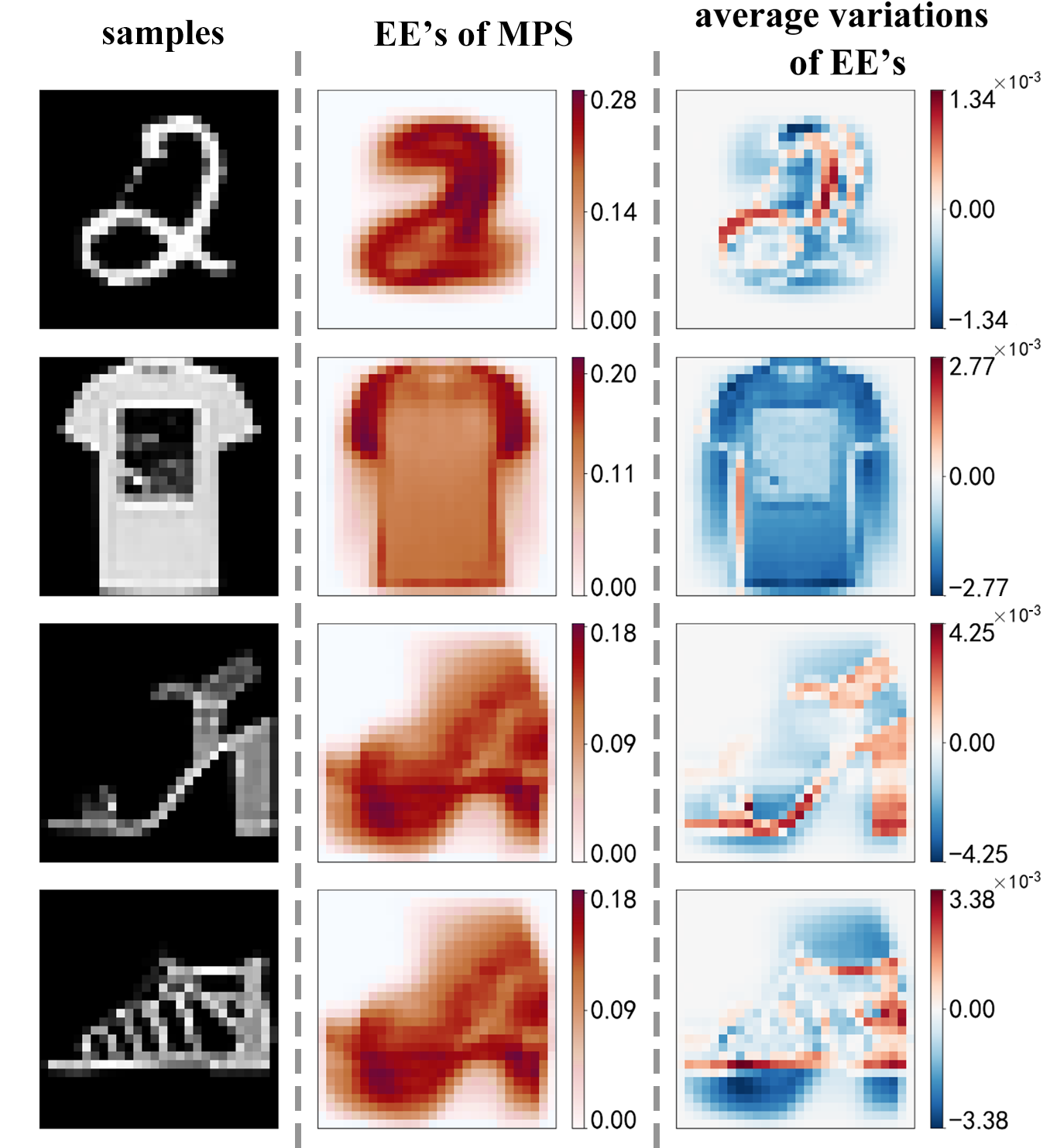}
    	\caption{(Color online) We take four training samples in the MNIST and fashion-MNIST datasets as the examples (first column). The EE of the generative MPS's $S$ [Eq.~(\ref{eq-EE})] and the variations of EE by measuring the generative MPS's [Eq.~(\ref{eq-dEE})] are demonstrated in the second and third columns, respectively.}
    	\label{fig-CM}
\end{figure}

Considering a specific sample $\tilde{\boldsymbol{x}}$ with knowing $\tilde{x}_{3} = 0$, we accordingly measure on the third physics index of $\boldsymbol{\Psi}$ in Eq.~(\ref{eq-singlet}) by following Eq.~(\ref{eq-Phik}), and have $\boldsymbol{\Phi}^{[3]} = [1, 0]^{T} \otimes [0, 1]^{T}$, with $S = 0$ for the rest two qubits. The average variations of EE $\langle \delta S \rangle_{1} = \langle \delta S \rangle_{2} = (0 - \ln{2}) / 2 = -(\ln{2})/2$ are negative. One can see that the last two qubits are highly entangled before the measurement, similar to the qubits in a same vertical strip. The measurement (in the basis of $[0, 1]^{T}$ and $[1, 0]^{T}$ in this case) on one qubit will generally eliminate the uncertainty of the other. This simple example implies that in the more complicated cases, a negative $\langle \delta S \rangle_{m'}$ might suggest a relatively large entanglement between the measured qubit and (some of) the rest. The measurement on such a qubit (in other words, in the condition of knowing the value of the corresponding feature) will result in a probability distribution with smaller uncertainty. Obviously, the same discussions can be made if we reverse the black and white colors in the images. 

\section{Testing on sophisticated datasets}

We test our scheme on more sophisticated datasets, which are the MNIST dataset with the images of hand-drawn digits, fashion-MNIST dataset with the images of articles, and the images of brain cells. For each class in a dataset, we train a generative MPS for evaluating the entanglement properties. We take four training samples as examples shown in the first column of Fig.~\ref{fig-CM}. The second column demonstrates the $S$ of the MPS's. The relatively large EE (red regions) marks the informative areas that approximately form the shapes of the corresponding digits or articles. Note again the informative areas are from the properties of the generative MPS's, thus do not depend on any specific samples. For instance, the last two sub-figures in the second column are the same, showing the EE of the generative MPS for shoes. 

The third column shows the average EE variations $\langle \delta S \rangle_{m'}$, which identify the critical monitories of the specific images shown in the first column. The distinct shapes of the original images are successfully outlined by $\langle \delta S \rangle_{m'}$. For instance, the special writing habit in the ``2'', the rectangle printed on the T-shirt, and the different styles of the shoes are reflected by $\langle \delta S \rangle_{m'}$, which do not appear in the illustrations of the $S$ of the MPS shown in the middle column. Particularly, the critical minorities of the two shoes are obtained from the same generative MPS, and the distinct shapes of these two images are still well identified by $\langle \delta S \rangle_{m'}$. 

An interesting observation is that the average variations can be positive or negative. It means the EE may increase after the measurement, differing from the toy dataset where the EE always decreases. The signs of the EE variations indicate the entanglement structure among the qubits of the MPS. As discussed above, a negative $\langle \delta S \rangle_{m'}$ suggests a relatively large entanglement between the measured qubit and some rest ones. To understand a positive $\langle \delta S \rangle_{m'}$, let us consider another simple example as
\begin{eqnarray}
  \boldsymbol{\Psi} = \frac{1}{\sqrt{3}} 
  \Big(
&&\left[ \begin{array}{cc}  
1 \\ 0 \end{array}  \right]
\otimes
  \left[ \begin{array}{cc}  
0 \\ 1 \end{array}  \right]
\otimes
\left[ \begin{array}{cc}  
1 \\ 0 \end{array}  \right]
+
\left[ \begin{array}{cc}  
0 \\ 1 \end{array}  \right]
\otimes
  \left[ \begin{array}{cc}  
1 \\ 0 \end{array}  \right]
\otimes
\left[ \begin{array}{cc}  
1 \\ 0 \end{array}  \right] 
\nonumber \\ &&+
\left[ \begin{array}{cc}  
0 \\ 1 \end{array}  \right]
\otimes
  \left[ \begin{array}{cc}  
0 \\ 1 \end{array}  \right]
\otimes
\left[ \begin{array}{cc}  
0 \\ 1 \end{array}  \right]
\Big)
  \label{eq-example2}
\end{eqnarray}
This describes the probability distribution of three samples $\boldsymbol{x}^{[1]} = (0, 1, 0)$,  $\boldsymbol{x}^{[2]} = (1, 0, 0)$, and $\boldsymbol{x}^{[2]} = (1, 1, 1)$ with identical probabilities. All the three qubits are entangled, where the EE of the first two qubits is $S \simeq 0.6365$. Consider again a sample $\tilde{\boldsymbol{x}}$ with $\tilde{x}_{3} = 0$. By measuring on the third qubit accordingly, the first two qubits will be projected into a maximally entangled state with $S = \ln{2} > 0.6365$. The EE increases after the measurement with $\langle \delta S \rangle > 0$. In this case, the probability of $P(x_{k}=\tilde{x}_{k})$ is in general small (note $x_{k}$ denotes the feature corresponding to the measured qubit and $\tilde{x}_{k}$ denotes the value of this feature in the specific sample). The measurement will (relatively) largely enhance the probabilities of the samples that also have $x_{k}=\tilde{x}_{k}$, which might be small before the measurement. Consequently, the uncertainty of the unmeasured qubits might increase, leading to $\langle \delta S \rangle > 0$.

\begin{figure}[tbp]
    	\centering
    	\includegraphics[angle=0,width=1\linewidth]{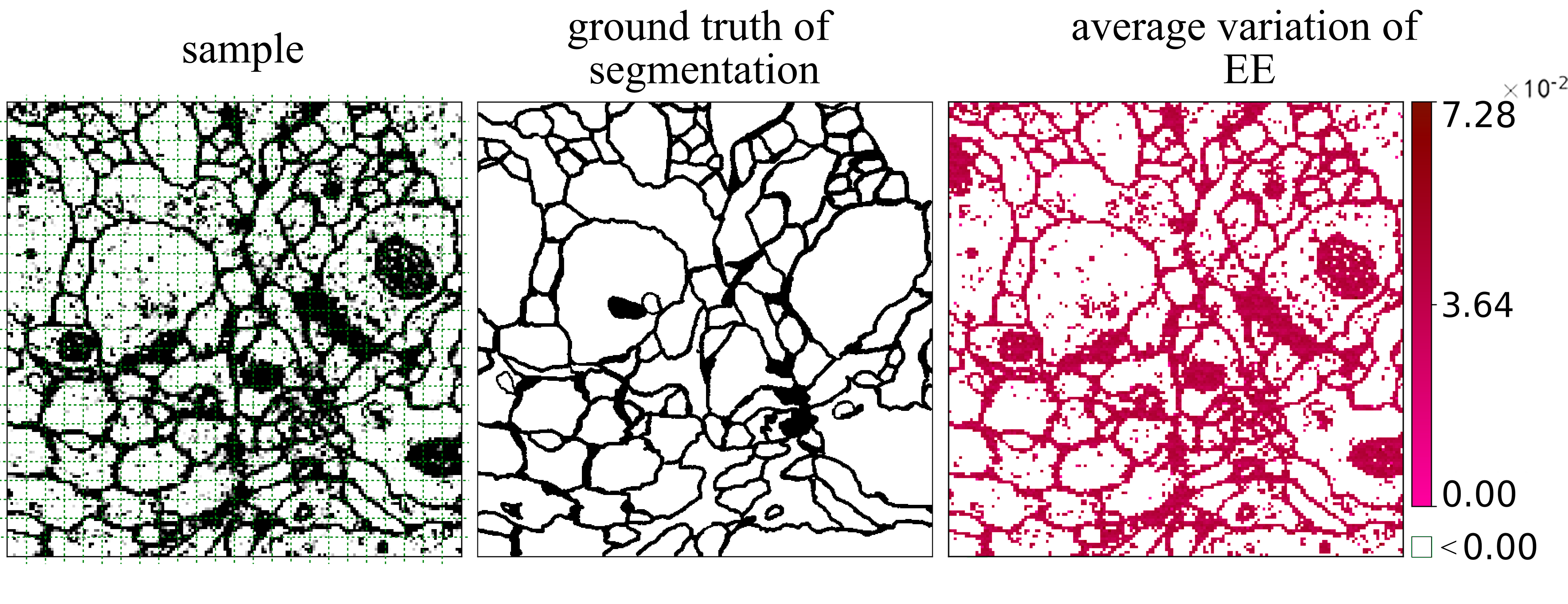}
    	\caption{(Color online) Image segmentation on the dataset of cell images~\cite{segCell}.}
    	\label{fig-vein}
\end{figure}

To further test our proposal, we apply our method on the images of brain cells for the purpose of unsupervised segmentation~\cite{segCell}. We take just one of the images and split it into ($24 \times 24$) pieces, of which each contains ($7 \times 7$) pixels. See the left subfigure of Fig.~\ref{fig-vein}. These pieces are subsequently fed to the TN as the training samples. The ground truth of the segmentation (i.e., the label) and the $\langle \delta S \rangle_{m'}$ given by the TN are shown in the middle and right subfigures, respectively. Note the $\langle \delta S \rangle_{m'}$ is obtained by joining together the results from all pieces. The negative $\langle \delta S \rangle_{m'}$ are all marked as white, since they are mainly the fluctuations caused by the boundary effects from the splitting. The cytoplasms are well separated from the cell membranes and nucleus. We shall stress that we take only one image from the dataset, and do not use any labeling information to obtain the right subfigure. Thus, our method belongs to the unsupervised segmentation schemes. The weakness is that our method cannot distinguish the nucleus from membranes. A possible solution is to introduce multi-qubit measurements instead of simple single-qubit measurements in the definition of $\langle \delta S \rangle_{m'}$. We provide more discussions and results in the appendix~\ref{app-noise} by testing on a toy dataset with noises. 

\section{Summary}
	In this work, we propose to utilize quantum entanglement for an unsupervised recognition of informative features. By training a generative tensor network (TN) that represents the probability distribution of features, measurements are used to transform the features, which are statistically meaningless, to the variations of the quantum entanglement entropy that identify the informative features. The proposed method is tested on a toy dataset of strips, MNIST dataset of hand-drawn digits, fashion-MNIST of articles, and the medical images of brain cells.  Our work sheds light on developing new interpretability schemes of machine learning via quantum information theories and TN methods, which can be applied to the unsupervised segmentation.

\appendix

\section{Variations of entanglement entropy by measurement}

\begin{figure}[tbp]
    	\centering
    	\includegraphics[angle=0,width=0.9\linewidth]{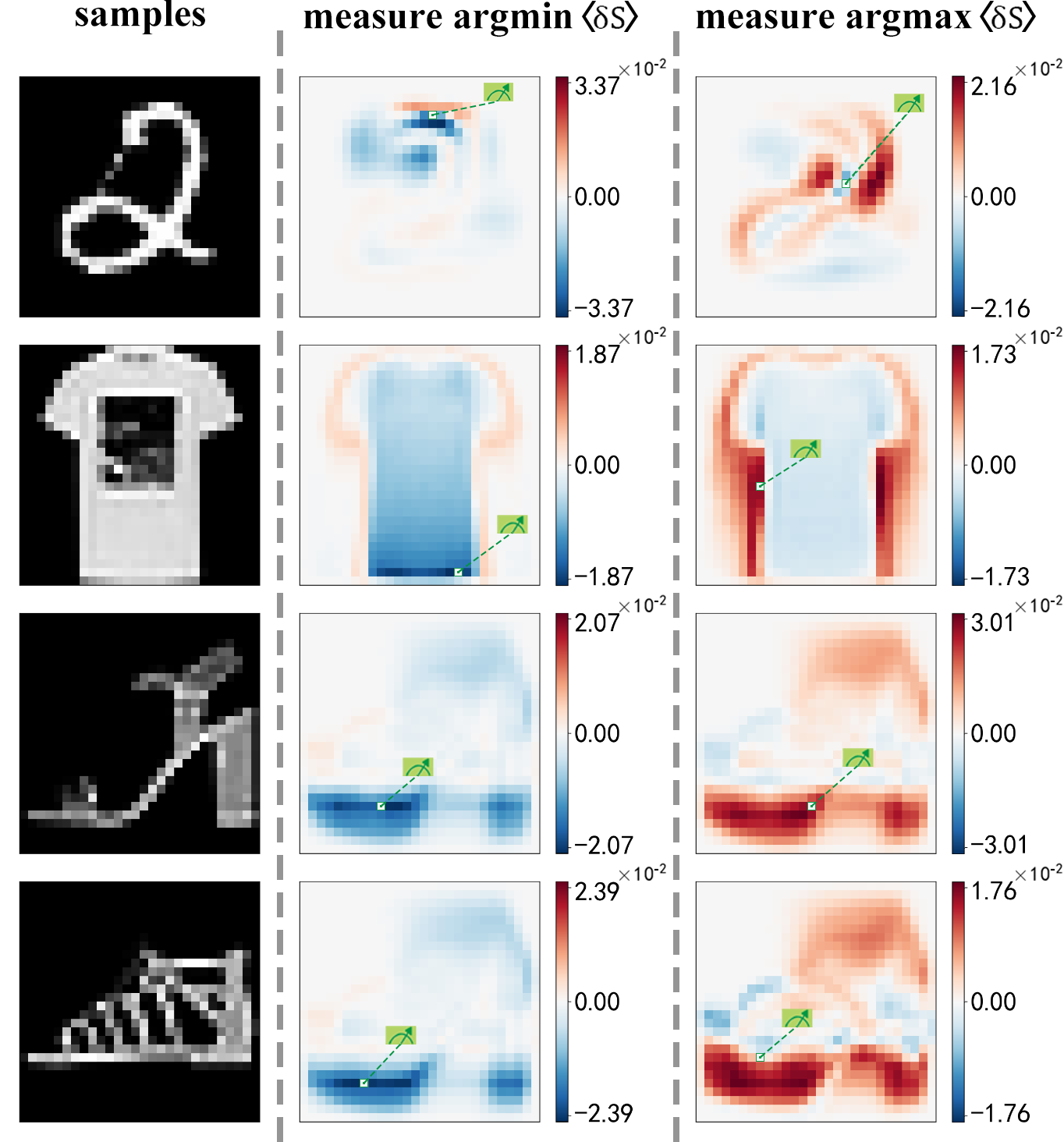}
    	\caption{(Color online) Four images from the MNIST and fashion-MNIST datasets (first column), and the variations of the entanglement entropy $dS_{m} = S'_{m} - S_{m}$ by taking $m = \arg\min_{m'} \langle \delta S \rangle_{m'}$ (second column) and $\arg\max_{m'} \langle \delta S \rangle_{m'}$ (third column). The green symbols with dash line show the positions of the measure qubits.}
    	\label{fig-measure}
\end{figure}

In Fig.~\ref{fig-measure}, we show how the entanglement entropy (EE) varies after measuring on a specific qubit. The first column shows four images taken from the MNIST and fashion-MNIST datasets as examples. We evaluate $\langle \delta S \rangle_{m'}$ ($m' = 0, \ldots, M$) by Eq.~(\ref{eq-dEE}) according to these images, respectively, and find the pixels with the algebraically smallest and largest variations $m_{\text{min}} = \arg\min_{m'} \langle \delta S \rangle_{m'}$ and $m_{\text{max}} = \arg\max_{m'} \langle \delta S \rangle_{m'}$. The corresponding variations of EE ($dS_{m} = S'_{m} - S_{m}$) are demonstrated in the second and third columns, respectively. The position of the measured qubit is marked by a green symbol with dash line. 

For the considered datasets, we have $\langle \delta S \rangle_{m_{\text{min}}} < 0$. Comparing with the EE of the MPS (see, e.g., Fig.~\ref{fig-CM} in the main text), the measurement on this qubit according to the value of the corresponding pixel will reduce the EE in the informative area. Remind that the informative area is defined by the pixels with large EE from the unmeasured MPS. Meanwhile, we also obverse certain positive variations that tend to locate at the edges of the informative area. By measuring on the $m_{\text{max}}$-th qubit, the qubits whose EE increases in general tend to locate at the edges of the informative area. 

\section{Benchmark on testing set}
\begin{figure}[tbp]
    	\centering
    	\includegraphics[angle=0,width=1\linewidth]{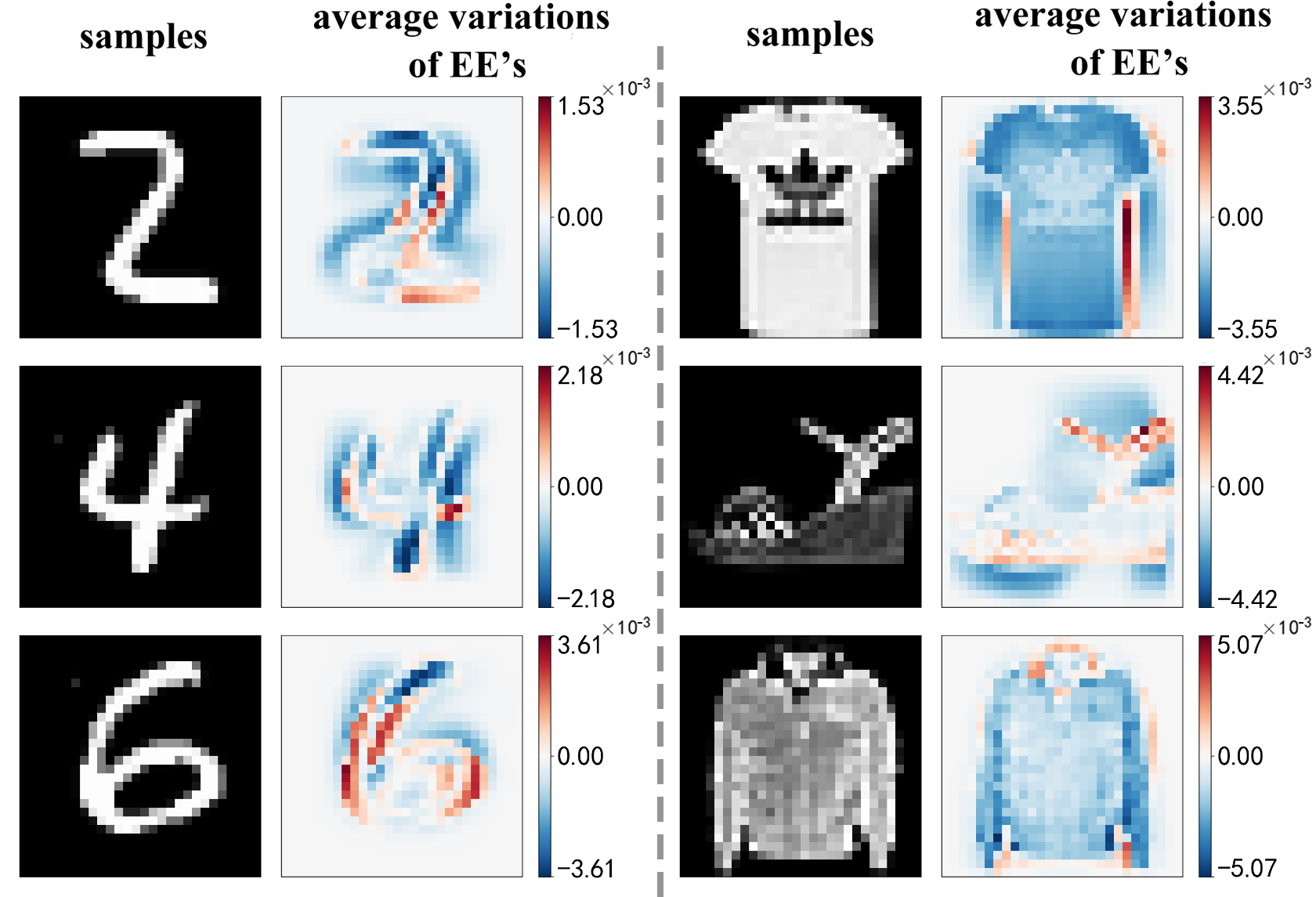}
    	\caption{(Color online) Six samples from the testing set of MNIST and fashion-MNIST datasets and the average variations of EE.}
    	\label{fig-test}
\end{figure}

Our scheme can be generalized to recognize the informative features and critical minorities of the samples that the generative TN has not learnt, e.g., those in the testing set. Fig.~\ref{fig-test} show six testing images from the MNIST and fashion-MNIST datasets as the examples to show the $\langle \delta S \rangle_{m'}$ by implementing measurements. Be aware that we do not have any prior information on labeling the pixels even for the training set. The $\langle \delta S \rangle_{m'}$ can be considered as the labels of the pixels that characterize importance to the given sample. By eyes one could recognize the distinct shapes in the testing images from $\langle \delta S \rangle_{m'}$.

\section{Robustness with different hyper-parameters}

The recognition of the informative features by the EE variations is robust to the changes the values of hyper-parameters. We generalize the feature map given in Eq.~(\ref{eq-featuremap}) to the following form, where the $s$-th element of the vector $\boldsymbol{v}$ from a given feature $x$ satisfies
\begin{eqnarray}
x \to v_{s} = \sqrt{\binom{d-1}{s-1}} \cos \left( \frac{\theta \pi}{2} x \right)^{d-s} \sin \left( \frac{\theta \pi}{2} x \right)^{s-1},
 \label{eq-GFM}
\end{eqnarray}
with $\binom{d-1}{s-1}$ the combination number and $\dim(\boldsymbol{v})=d$ that is also the dimension of the physical indexes of the MPS. By taking $d=2$ and $\theta=0.5$, Eq.~(\ref{eq-GFM}) is reduced to the feature map in Eq.~(\ref{eq-featuremap}). We choose $\theta=0.5$ in the main text since the classifier formed by the generative MPS's gives the highest classification accuracy~\cite{SPLRS19GTNC}. 

\begin{figure}[tbp]
	\centering
	\includegraphics[angle=0,width=1\linewidth]{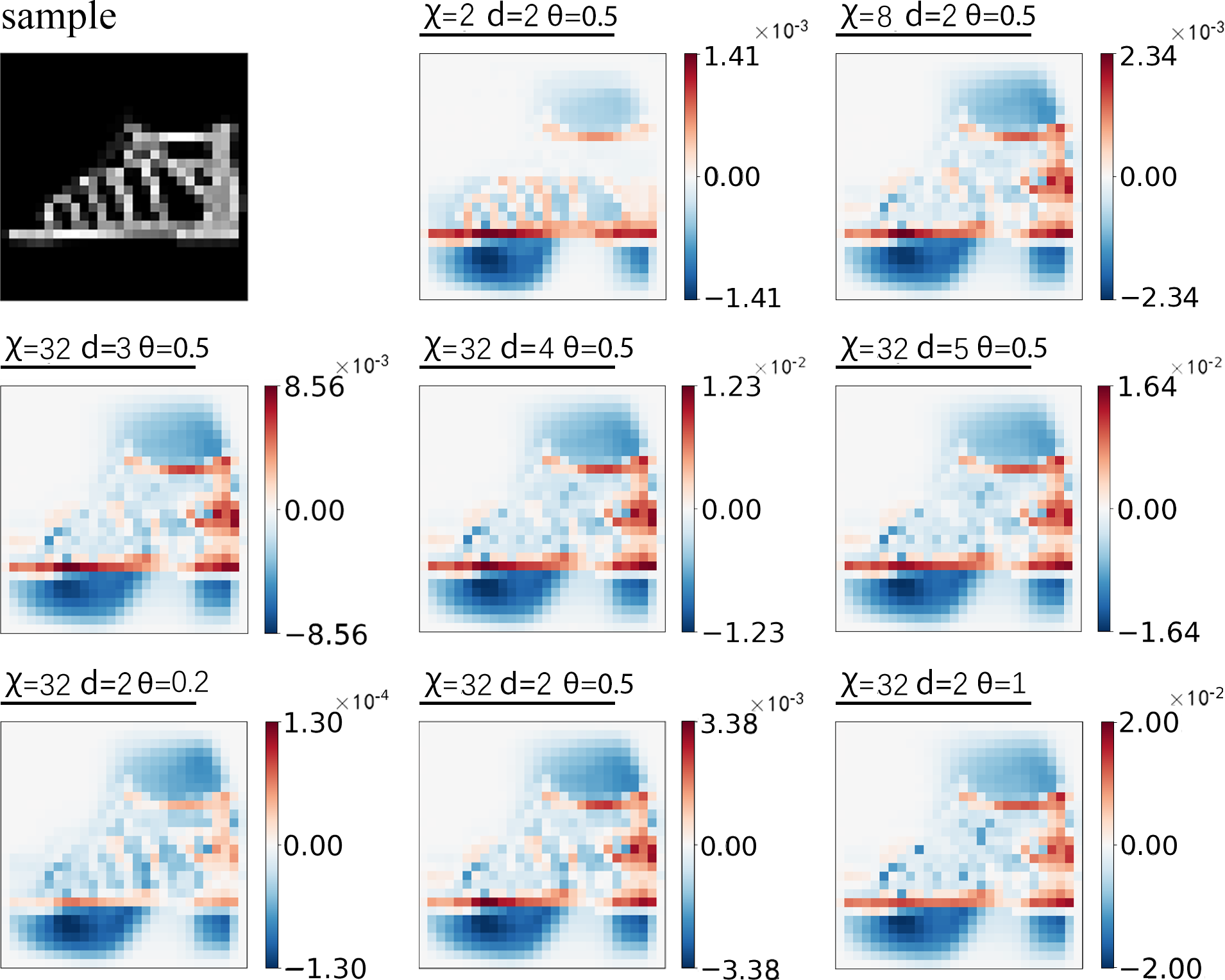}
	\caption{(Color online) We take one specific image of shoe as an example (top-left) and show the variations of EE with different values of $\chi$, $d$, and $\theta$.}
	\label{fig-hyperpara}
\end{figure}

In Fig.~\ref{fig-hyperpara}, we show the average variations of EE ($\langle \delta S \rangle_{m'}$) from a same image of shoe by taking different values of $\chi$ (the dimension of the virtual indexes), $d$ and $\theta$. In all cases, the distinct shapes at different parts of the shoe is well presented by $\langle \delta S \rangle_{m'}$ with slight differences. For instance , the strips of the shoe can be seen clearly with $\chi=2$, $d=2$, and $\theta=0.5$ (top-middle of Fig.~\ref{fig-hyperpara}), and the back counter is clearly captured with $\chi=32$, $d=2$, and $\theta=1$. In all cases, the sole and the topline are well presented. Ba aware that such distinct shapes of this specific image cannot be seen by the EE of the MPS, which is sample independent.

\section{Entanglement entropy for feature selection}

\begin{figure}[tbp]
	\centering
	\includegraphics[angle=0,width=1\linewidth]{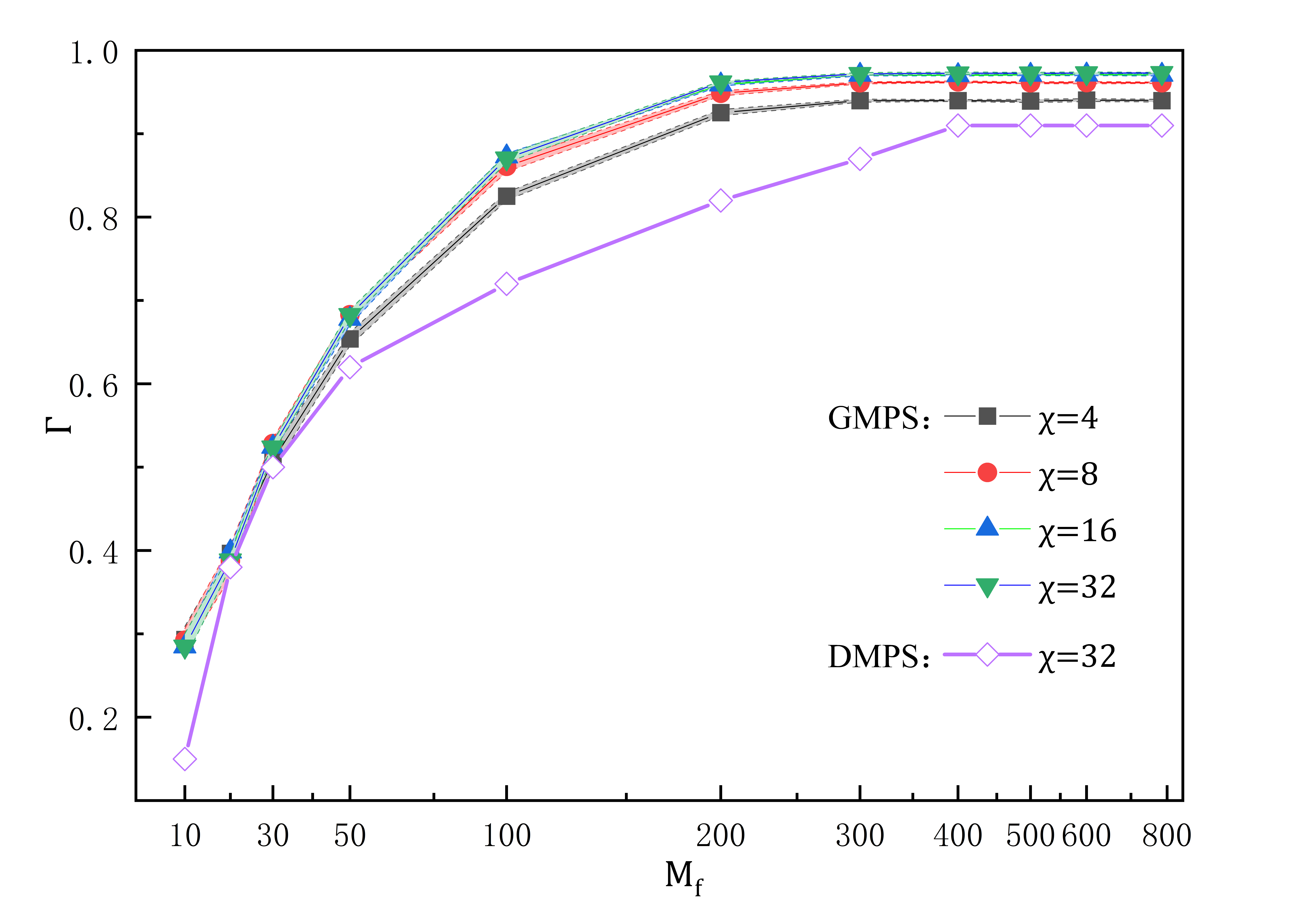}
	\caption{(Color online) Testing accuracy $\Gamma$ versus the number of features $M_{f}$ selected according to the EE. The solid symbols with lines show the accuracy of the generative MPS's with $\chi=4$, $8$, $16$, and $32$. The thickness of the lines illustrates the variance evaluated by repeating the simulations for ten times. The purple hollow diamonds show the accuracy of the feature selection approach based on the discriminative MPS approach~\cite{LZLR18entTNML}.}
	\label{fig-acc}
\end{figure}

The EE of the MPS can be utilized for feature selection. Following the idea of the generative MPS classification scheme proposed in Ref.~[\onlinecite{SPLRS19GTNC}], we take the MNIST as example and train a generative MPS for each of the classes. The classification is implemented by comparing the fidelity (a measure of similarity between two quantum states) between the product state [Eq.~(\ref{eq-featuremap})] from a target sample and the MPS's. For each MPS, we retain $M_{f} <M$ (note $M$ is the total number of features in one sample) features whose qubits possess the largest EE. Fig.~\ref{fig-acc} shows the testing accuracy versus $M_{f}$ with $\chi=4$, $8$, $16$, and $32$. By repeating ten times of simulations with a randomly initialized MPS, the thickness of the lines shows the variances that are insignificant. 

Our accuracy surpasses that of the feature selection method proposed in Ref.~[\onlinecite{LZLR18entTNML}] that is based on the EE of the discriminative MPS~\cite{SS16TNML}. In the discriminative MPS approach, one trains an MPS that contains $M$ physical indexes and an additional $D$-dimensional label index (with $D$ the number of classes). The $M_{f}$ features with the largest EE evaluated from the discriminative MPS are retained. Our method with the generative MPS classification scheme demonstrated obvious advantage for about $100 < M_{f} < 300$ and for the few-shot cases around $M_{f} \simeq 10$.

\section{Generalizing to multi-qubit measurements to deal with single-pixel noises}
\label{app-noise}

\begin{figure}[tbp]
	\centering
	\includegraphics[angle=0,width=1\linewidth]{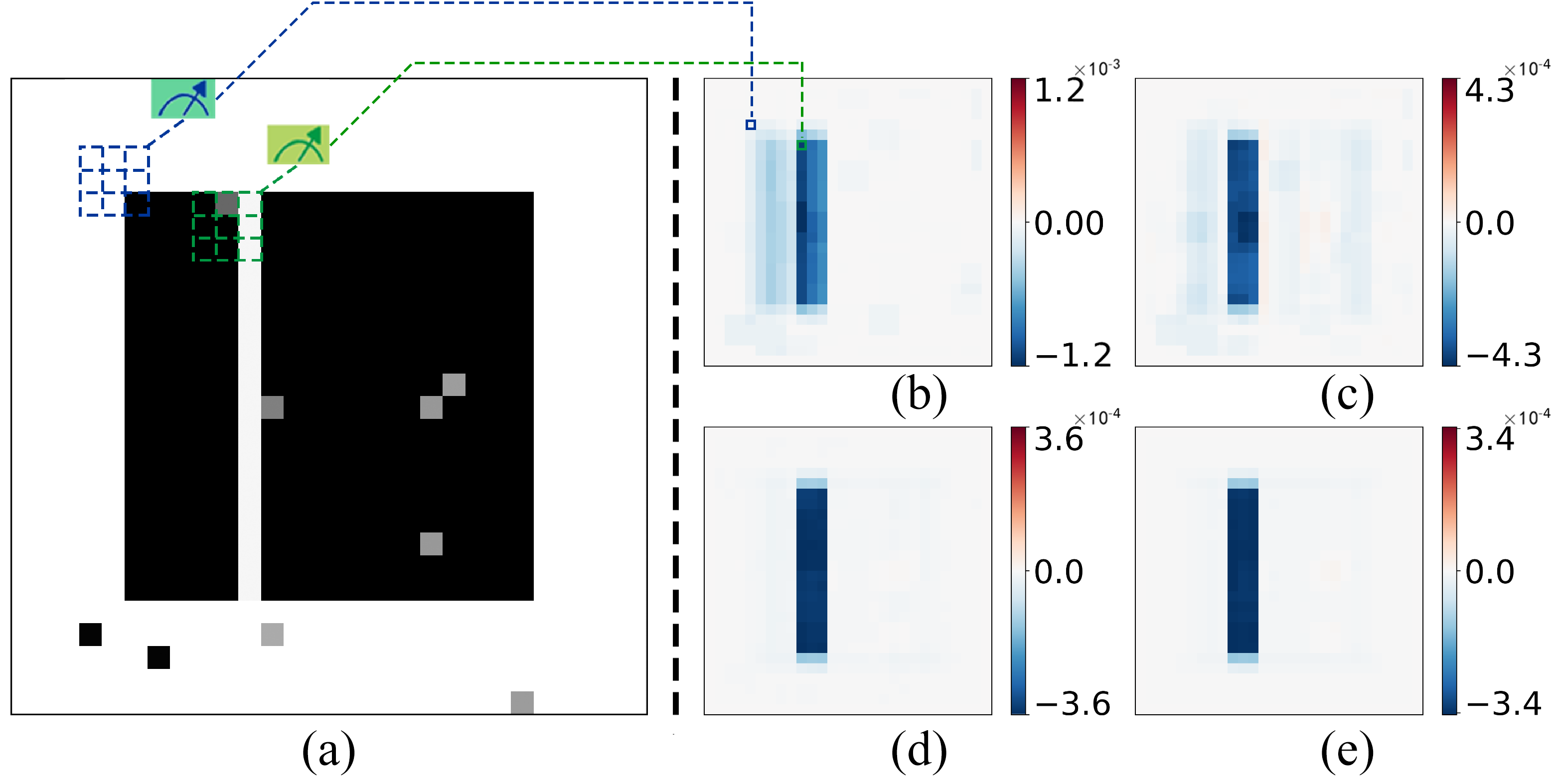}
	\caption{(Color online) (a) An example in the noisy strip dataset. (b)-(e) The average variation of EE $\langle \delta S \rangle_{m'}$ with different numbers ($6$, $18$, $180$, and $540$, respectively) of noisy training samples. The noises can be better excluded from the critical minority with more training samples.}
	\label{fig-noise}
\end{figure}

From the definition of the EE variations, the EE of each qubit is solely determined by the value of the corresponding pixel and the MPS, thus mathematically cannot distinguish the noises. In Fig.~\ref{fig-noise}, we show by the multi-qubit measurements, the single-pixel noises can be excluded from the critical minority. Fig.~\ref{fig-noise} (a) shows an example of the noisy strip dataset. For each time, we measure ($3 \times 3$) neighboring qubits demonstrated by the blue dash squares. The average variations of EE $\langle \delta S \rangle_{m'}$ after the measurement are shown in Fig.~\ref{fig-noise} (b)-(e), where $m'$ denotes the numbering of the ($3 \times 3$) square for the measurement. The MPS is trained by $6$, $18$, $180$, and $540$ noisy samples, respectively. When the number of training samples is small, the noisy pixels in the background can be well excluded. But the $\langle \delta S \rangle_{m'}$ in the informative area show certain ``fluctuations''. When the number of training samples increases, our method can better select the informative features in all areas. 

\section*{Acknowledgment} We are indebted to Yuhan Liu, Zheng-Zhi Sun,  Ke Li, Peng-Fei Zhou, Rui Hong, Jia-Hao Wang, and Wei-Ming Li for helpful discussions. This work was supported by NSFC (Grant No. 12004266 and No. 11834014) and Foundation of Beijing Education Committees (No. KM202010028013). SJR acknowledges the support from the Academy for Multidisciplinary Studies, Capital Normal University.



\section*{Competing Interests}
The authors declare no competing financial or non-financial interests.

\section*{Data Availability}
The authors will provide the codes and data under reasonable requests. 

\section*{Author Contributions}
Shi-Ju Ran conceived the idea of this work; all authors contributed to the codes;  Sheng-Chen Bai and Yi-Cheng Tang implemented the numerical simulations; all authors contributed to the writing of the manuscript. 

\normalem
%


\begin{thebibliography}{36}
\bibitem [{not()}]{note1}%
  \BibitemOpen
  \href@noop {} {}\bibinfo {note} {Here we avoid to use Dirac's symbols for the
  readers who are not familiar with quantum physics. We use a bold letter to
  represent a tensor (or matrix, vector, such as $\boldsymbol{\Psi}$), and the
  same normal letter with lower indexes to represent the tensor elements (such
  as $\Psi_{s_{1} \ldots s_{M}}$).}\BibitemShut {Stop}%
\end{thebibliography}

\end{document}